\documentclass[twocolumn]{aastex61}

\usepackage{amsmath, amsthm, amssymb}
\usepackage{bm}
\usepackage{multirow}
\usepackage{xcolor}                                                                      
\usepackage[T1]{fontenc}
\usepackage{colortbl} 

\shorttitle{Tidal disruption of proto-Mercury}
\shortauthors{Deng}

\begin{document}

\title{hypothetical hyperbolic encounters between Venus and proto-Mercury that partially stripped away proto-Mercury's mantle}
\correspondingauthor{Hongping Deng}
\email{hd353@cam.ac.uk}

\author{Hongping Deng}
\affiliation{Department of Applied Mathematics and Theoretical Physics, University of Cambridge, Centre for Mathematical Sciences, Wilberforce Road, Cambridge CB3 0WA, UK}
\affiliation{Center for Theoretical Astrophysics and Cosmology, Institute for Computational Science, University of Zurich, Winterthurerstrasse 190, 8057 Zurich, Switzerland}

\begin{abstract}
Mercury has an unusually large metal core comprising $\sim$70\% of its mass comparing to all other terrestrial planets in the solar system. Giant impacts can remove a significant fraction of the silicate mantle of a chondritic proto-Mercury and form the iron rich present-day Mercury. However, such high-temperature giant impacts seem at odds with the retainment of moderately volatile elements on present-day Mercury \citep{Peplowski2011}. We simulated a series of hyperbolic encounters between proto-Mercury and proto-Venus, which may occur in the chaotic early solar system. Tidal disruption of  proto-Mercury always removes part of its silicate mantle while its iron core remains intact. We find, in favourable cases, four close encounters with fast spinning projectiles (resulting from previous encounters) can lead to present-day Mercury iron fraction. More encounters are needed when the spin and orbital angular momentum are not always aligned. These hyperbolic encounters have various outcomes, such as orbital decay, binary planets and change of spin rates. These results suggest the importance of proper treatment of close encounters in N-body simulations of planetary accretion. 

\end{abstract}
\section{Introduction}
The four terrestrial planets in our solar system, Mercury, Venus, Earth and Mars are believed to form through collisional growth between planetesimals. Mercury, the smallest one, poses a serious challenge to N-body simulations of terrestrial planet formation \citep[see, e.g.,][and references therein]{Lykawka2017, Clement2019}. Mercury also stands out for its unusually high mean density \citep{Anderson1987} which suggests a high iron mass fraction $Z_{Fe}\sim 70\%$ \citep{Hauck2013}. However, all the other planets have $Z_{Fe}\sim 30\%$, in accordance with the protosolar iron abundance. Giant impacts onto a roughly chondritic proto-Mercury ($Z_{Fe}=30\%$) can remove a significant fraction of proto-Mercury's silicate mantle and explain the high iron mass fraction \citep{Benz1988, Benz2007,Asphaug2014, Chau2018}. Alternatively, iron/silicate fractionation during the planetary growth stage is also proposed to explain the high $Z_{Fe}$ in Mercury \citep[see review by][]{Benz2007, Ebel2017}.

There are two classes of giant impacts proposed. Proto-Mercury may experience one (or possibly more) impact by a less massive projectile \citep{Benz2007, Chau2018} or itself hits proto-Venus or proto-Earth and manages to run away \citep{Asphaug2014, Chau2018}. A collision between two protoplanets seems artificial, but collisional trajectories do exist even for present-day Mercury and Venus \citep{Batygin2008, Laskar2009}. Indeed, collision/encounters between proto-Mercury and proto-Venus are observed in N-body simulations of terrestrial planet formation \citep[e.g.][]{Hansen2009}. The true challenge for the giant impact hypothesis comes from the results of the MESSENGER mission where a relatively high abundance of volatile elements, such as potassium (K), thorium(Th) and uranium (U) are measured excluding formation scenarios of Mercury involving a high-temperature process \citep{Peplowski2011}. These observational constrains lead us to wonder if there are low-temperature (without intensive shock heating) mechanisms that can remove proto-Mercury's mantle significantly. 

Tidal disruption of a minor body can happen when it passes close by proto-Venus or proto-Earth. Such close encounters should be more likely than direct collisions onto protoplanets but remain unexplored \citep[but see,][]{Asphaug2006}. What would happen should proto-Mercury experiences a close encounter with proto-Venus? Could tidal disruption strip away proto-Mercury's mantle and explain the high $Z_{Fe}$? We carried out direct hydrodynamical simulations of close encounters  between proto-Mercury and proto-Venus with the messless finite mass (MFM) method in the GIZMO code \citep{Hopkins2015,Deng2019a}. We aim at answering the above questions and exploring the outcomes of such close encounters. We first describe how we set up hyperbolic encounters and the MFM method in section \ref{sec:model}. The evolution and analysis of close encounters are reported in section \ref{sec:results} and conclusion follows in section \ref{sec:con}.
\section{Models and simulations}
\label{sec:model}
\subsection{Encounter geometry}
\label{sec:hyperbolic}
\begin{figure}[ht!]
\epsscale{1.2}
  \plotone{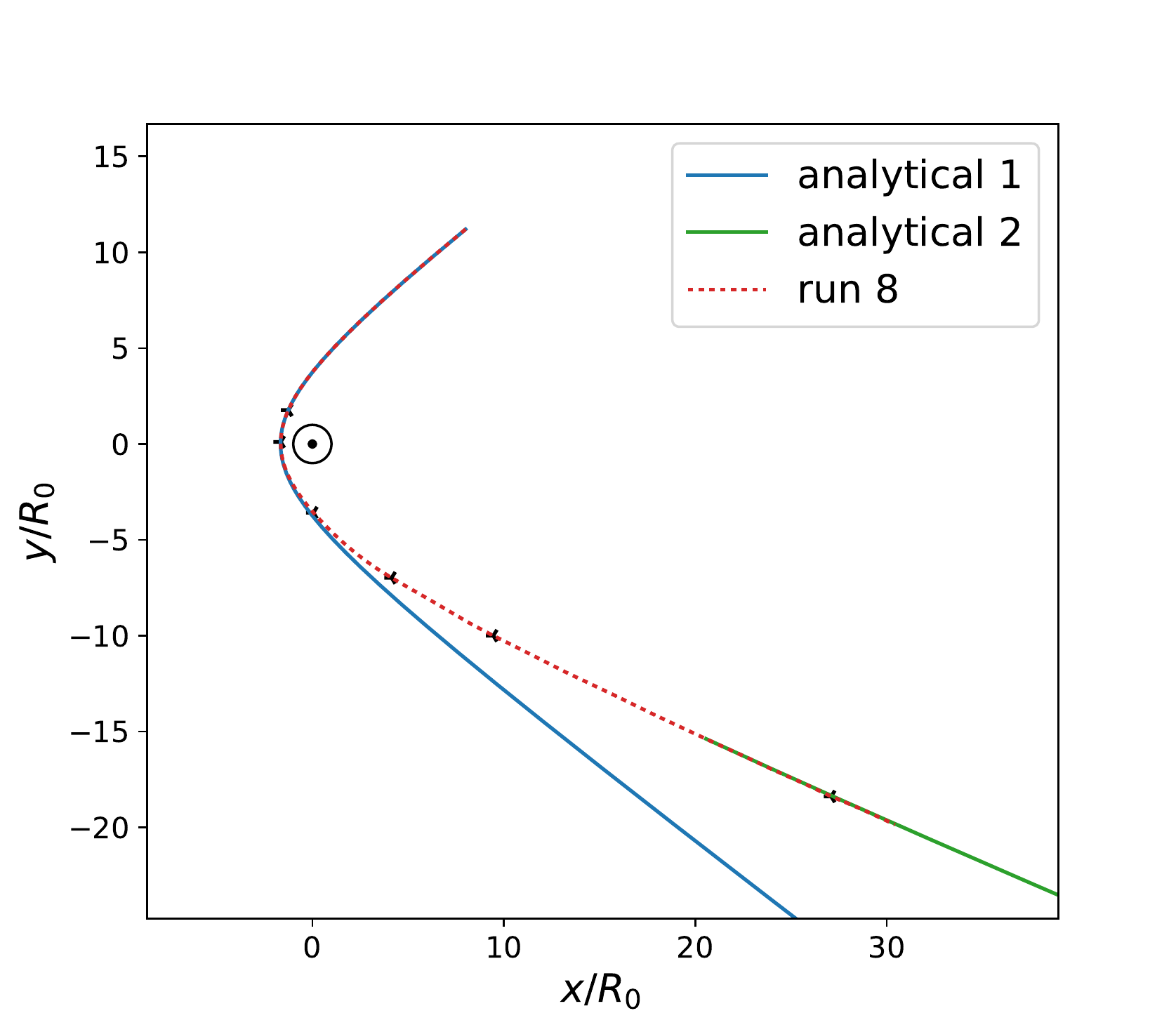}
\caption{The prescribed (blue curve) hyperbolic, post-encounter (green curve) hyperbolic and the real (red curve) orbits of run 8 in table \ref{tab:simulations}. The blue and green curves are analytical solutions of the two-body Kepler problem showing the displacement vector while the red dashed line indicates the relative position of the center of mass of the target's and projectile's cores. The simulation follows the prescribed hyperbolic orbit (blue curve) well before losing orbital energy due to tidal interaction and entering a lower energy hyperbolic orbit (green curve). The six markers indicate where we take snapshots for analysis in figure \ref{fig:run8}. The black circle denotes the radius of Venus, $R_0=1$. \label{fig:encounter}}
\end{figure}

When two protoplanets are far from each other their tidal interaction can be neglected so that their interaction reduce to the standard two-body Kepler problem. Here we focus on hyperbolic orbits in the Kepler problem. Such hyperbolic orbits (see e.g., figure \ref{fig:encounter} blue curve) can be characterized by two quantities: the closest approach, $d$ and their relative velocity at infinity $v_\infty$. To setup the trajectories we need to work out the hyperbolic orbits with specified $d$ and $v_\infty$ in Cartesian coordinates. A standard hyperbola reads,

\begin{equation}
\frac{x^2}{a^2}-\frac{y^2}{b^2}=1, \label{eq:hyperbola}
\end{equation}
with $a, c=\sqrt{a^2+b^2}, e=c/a$ being the vertex, focus and eccentricity. The maximum relative velocity $v_{max}$ occurs at the vertex corresponding to the closest approach, $d=c-a=a(e-1)$. The specific energy conservation requires
\begin{equation}
\frac{1}{2}v^2_{max}-\frac{1}{2}v^2_{\infty}=\frac{GM}{d}, \label{eq:energy}
\end{equation}
where $M=m_1+m_2$ is the sum of the masses.
The specific angular momentum conservation leads to,
\begin{equation}
dv_{max}=bv_{\infty},
\end{equation}
which is equivalent to $v_{max}=\sqrt{(1+e)/(1-e)}v_\infty$. If one substitutes it into equation \ref{eq:energy} one finds,
\begin{equation}
a^2=\frac{GM}{v^2_{\infty}}.
\end{equation}
We get $e=d/a+1$ and all other parameters for the hyperbola by definition. 

In our simulations we always set an initial $x$ coordinate offset, $\delta x_0=8R_0$ with $R_0$ denoting the radius of Venus. $\delta y_0$ can be calculated from equation \ref{eq:hyperbola} and the velocity at this point is calculated similarly to equation \ref{eq:energy} at the vertex. Note the slope of the hyperbola, i.e., the velocity direction, is $dy/dx=b^2(\delta x_0+c)/a^2 \delta y_0$. We place the center of masses at the origin and at rest. At the beginning of the simulations, the tidal force is less than 0.1\% of its peak value at the closest approach. Indeed, the simulations follow the prescribed hyperbolic orbit well in the early stages (figure \ref{fig:encounter}). Near the end of the encounter simulations, we calculate the relative position and velocity between the two largest bodies (see section \ref{sec:results}).  The new relative velocity at infinity, $v^{'}_\infty$ can be calculated similarly to equation \ref{eq:energy} at vertex. We introduce a negative sign to $v^{'}_{\infty}$ when the two largest bodies are gravitationally bound (see table \ref{tab:simulations}).

\subsection{Encounter simulations}

We simulate the hyperbolic encounters using the meshless finite mass (MFM) method in the GIZMO code \citep{Hopkins2015}. The MFM method is a hybrid Lagrangian method different from the traditional Smoothed Particle Hydrodynamics (SPH) method \citep{Lucy1977, Gingold1977};  MFM does effective volume partition according to the particle distribution and then solves the Riemann problem to update the fluid variables and can be regarded as a generalized moving mesh method. It employs no explicit artificial viscosity and thus shows better conservation property than SPH \citep[see, e.g.,][]{Deng2017}. We added equation of state (EOS) library interfaces for both the ANEOS \citep{Thompson1974} and Tillotson EOS \citep{Tillotson1962} library to the GIZMO code in \citet{Deng2019a} and the ANEOS EOS is applied here in the encounter simulations \footnote{The GIZMO code with Tillotson EOS is publicly  available at \href{http://www.tapir.caltech.edu/~phopkins/Site/GIZMO.html}{\url{http://www.tapir.caltech.edu/~phopkins/Site/GIZMO.html}}}. In giant impact simulations,  MFM captures subsonic turbulence and thus mixing more accurately than traditional SPH \citep[see][and references therein]{Deng2019b, Deng2019a}. 

We build low noise proto-Mercury and proto-Venus (present-day Venus mass assumed) models based on equal area tessellations of the sphere detailed in \citet{Reinhardt2017} similarly to \citet{Deng2019b, Deng2019a}. Note here we have proto-Mercury models with different $Z_{Fe}$ but the same core mass which equals present-day Mercury core mass. Both proto-Venus and proto-Mercury are fully differentiated with an isentropic iron core and isentropic dunite mantle. The temperature in the center of the core is 4000K for proto-Venus and 1500K for proto-Mercury. To study the effects of proto-Mercury spin on the encounter, we spin up a non-spinning proto-Mercury gradually to the desired initial spin and relax the initial condition to an equilibrium state. The planetary bodies are then placed on hyperbolic orbits characterized by the closest approach and velocity at infinity (see section \ref{sec:hyperbolic}). We use about 500K particles in our encounter simulations, which is comparable to present-day high-resolution impact simulations. We aim at modelling the gravitational interaction instead of complex turbulence or the post-encounter thermal state \citep[see resolution study in][]{Deng2019b} so the resolution employed here should be sufficient. We use 1 Venus radius ($R_0$), 1 km/s with the gravitational constant equals 1 as our unit system. All simulations and key results are summarised in table \ref{tab:simulations}. 



  \section{Results}
\label{sec:results}
  \begin{figure*}[ht!]
\plotone{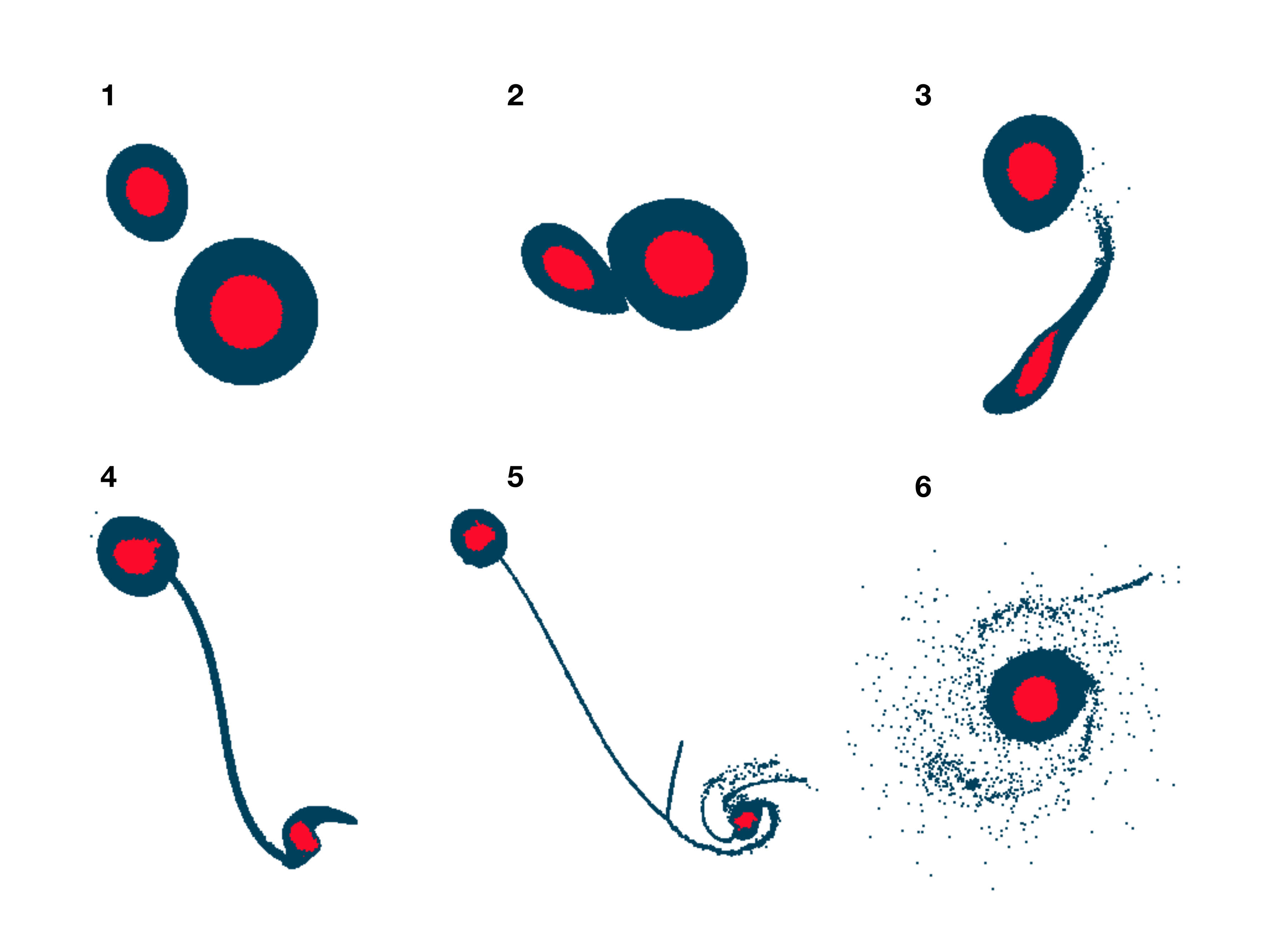}
\caption{Cutaway snapshots ($z<0$) of run 8 taken at t=-0.35h, 0, 1.00h, 2.87h, 5.70h, 17.38h as indicated in figure \ref{fig:encounter}. The snapshots show different components of the target and projectile with the red color indicating the iron cores. Note the last snapshot focuses on the escaping projectile and its disk. \label{fig:run8}}
\end{figure*}

We start describing the encounters between proto-Venus and a non-spinning chondritic proto-Mercury ($Z_{Fe}=30\%, R=0.59R_0$), i.e., run 1-12 in table \ref{tab:simulations}. Here we take run 8 as an example again. As shown in figure \ref{fig:encounter}, the projectile follows the prescribed hyperbolic orbit initially because the separation is larger than $10R_0$, and the tidal force is negligible. Figure \ref{fig:run8} shows the tidal interaction between the projectile and target of run 8 at positions marked on figure \ref{fig:encounter} (note the final snapshot zooms in on the projectile). When the separation is about $2R_0$, the projectile shows significant deformation to a droplet shape. At the closest approach, the projectile just passes the surface of the target in run 8. It may scratch or fly over the target's surface in lower or higher energy encounters. The projectile passes the closest approach largely intact, and the near side mantle is stretched strongly towards the target (t=1 h). The near side forms an elongated stream of silicates plunging deep into the target's mantle reaching its core-mantle boundary. The projectile has gained significant spin by now, and its far side forms a spiral due to  differential rotation (t=2.87 h). The spiral breaks up and collides further with the thin silicate stream. Eventually, all the pieces of the spirals are disrupted by the tidal force of the escaped proto-Mercury forming a debris disk. We stop the simulation after the debris disk forms (typically $t>20 $h, $\delta x>20R_0$).

We use the clump finder SKID \citep{Stadel2001} to identify the largest and second-largest bodies bound by gravity \citep[see also][]{Chau2018}.  The post-encounter proto-Mercury (the second-largest body) restores spherical shape with an equatorial radius that is slightly ($< 5\%$) larger than the polar radius. We calculate its total spin angular momentum and assign it a spin rate (table \ref{tab:simulations}) assuming rigid body rotation despite that the outer part rotates at a higher angular speed. The circumplanetary disk consists of particles gravitational bound to the second largest body determined following an established approach \citep[see, e.g.][]{Canup2013}. The disk mass is $<4\%$ of the post-encounter proto-Mercury. 

Table \ref{tab:simulations} shows one single hyperbolic encounter between a chondritic proto-Mercury and proto-Venus can at best promote $Z_{Fe}$ to $\sim35\%$ even when proto-Mercury is tidally captured (e.g., run 2). We carried out further simulations with increased initial iron mass fraction, $Z_{Fe}=40\%, 50\%, 60\%$. All these simulations are qualitatively similar, but the increment of $Z_{Fe}$ seems larger when the initial $Z_{Fe}$ is larger. The similar behaviours across a wide range of initial $Z_{Fe}$ suggest a series of encounters may increase $Z_{Fe}$ from $30\%$ to $70\%$ consecutively. Indeed, possible encounters between Mercury and Venus due to a dynamical instability do happen as a series of encounter events \citep[see, e.g.,][Fig 10]{Batygin2008}. We expect the removal of previous circumplanetary disks in following encounters due to the strong tidal force. 

The non-spinning proto-Mercury spins up significantly with a typical spin period of 4-6 hours (table \ref{tab:simulations}). This leads to the question of how the initial spin (gained from previous encounters or during proto-Mercury formation) affects the tidal disruption. To this end, we carried out encounter simulations with different initial spins (close to the spin gained from one encounter) for the $Z_{Fe}=40\%$ model. We expect similar results for various initial $Z_{Fe}$ as in the non-spinning simulations. When the initial spin is aligned with the orbital angular momentum (run 16-19), the efficiency of mantle removal increases with spin. $Z_{Fe}$ can increase $>10\%$ after one single encounter.
However, when the initial spin and orbital angular momentum is anti-aligned (run 20-23), the projectile is very resistive to tidal disruption (run 20, 21 are not disrupted at all). The strong tidal force during the close encounter manages to align the spin of the projectile with the orbital angular momentum except in run 20. 
\begin{deluxetable}{c|c|c|c|c|c|c|c|c}[ht!]
  \tablenum{1}
  \tablecaption{Hyperbolic encounter simulations\label{tab:simulations}}
  \tablewidth{\textwidth}
  \tablehead{
    \colhead{Run}  & \colhead{$Z_{Fe}$} & \colhead{$T_{spin}$} &\colhead{$v_\infty$} & \colhead{$d/R_{0}$}  & \colhead{$Z^{'}_{Fe}$} & \colhead{$T^{'}_{spin}$} & \colhead{$v^{'}_{\infty}$}&\colhead{$\frac{M_{disk}}{M_{2nd}}$}}
  \startdata
    1                            &30    & NA &  2.       & 1.5              & 33.3            & 4.05            &  -2.13              &  3.0              \\
2                             &30     & NA &  2.       &1.75               & 36.0             & 4.01            &  -1.21             &  3.5       \\
3                             &30     & NA &  2.       &2                & 33.1             & 4.60           &  0.58            &  1.7        \\
  4                            &30    & NA &  2.5       & 1.5              & 33.3            & 4.10            &  -1.36              &  3.7              \\
5                             &30     & NA &  2.5       &1.75               & 35.6             & 4.12            &  0.97             &  4.0         \\
6                             &30     & NA &  2.5       &2                & 32.2             & 5.40            &  1.58             &  1.2        \\
7                              &30     & NA  & 3       &1.5               &33.0              &4.52            & 1.07                & 3.5           \\
8                             &30     & NA & 3       &1.75               &34.7              &4.03            & 1.85                & 3.0            \\
  9                             &30     & NA  & 3      &2.              &31.7             &5.02            & 2.27               & 1.3                     \\
  10                           &30    & NA  & 3.5       &1.5                &32.7              &4.48           & 2.17                & 3.2                   \\
  11                           &30     & NA & 3.5       &1.75              &34.2             &5.07             & 2.57                & 3.3            \\
  12                           &30     & NA  & 3.5       & 2.0              &31.5              &5.25             &2.84                 & 1.6                 \\
13                             &40     & NA  & 3      & 1.5              &44.6              &4.30             & 1.48                &      4.2    \\
14                             &40     & NA  & 3      & 1.75              &46.1             &4.05            &2.17                 & 4.0        \\
15                            &40     & NA  & 3      & 2             &42.3              &4.74            &2.44              & 1.6         \\
16                               &40     & 3  & 3      & 1.75              &53.2              &2.69           &2.89                 & 6.8          \\
17                             &40     & 4  & 3      & 1.75              &49.1             &2.88            &2.67                 & 3.1           \\
18                            &40     & 5  & 3      & 1.75              &48.6              &3.37             &2.6                 & 4.4         \\
19                            &40     & 6  & 3      & 1.75              &47.8             &4.39            &  2.55              & 4.5         \\
20                            &40      & -3  & 3      & 1.75              & 40             &-20.26             &2.45                 & 0              \\
  21                           &40    & -4  & 3       &1.75                &  40              &  7.67                         &  2.19              & 0         \\
  22                           &40     & -5 & 3       &1.75              &    40.5                   &   5.4                      &2.03                &1.3             \\
  23                           &40     & -6  & 3       & 1.75              &41.2             &4.62            &1.93                 & 1.6              \\
   24                           &50    & NA  & 3       &1.5                &56.8              &6.83            & 1.67                & 7.1             \\
  25                           &50     & NA & 3       &1.75              &56.6              &4.55           & 2.33               & 3.7                \\
  26                           &50     & NA  & 3       & 1.75             &52.0              &5.13             &2.52                & 1.7                  \\
  27                           &60    & NA  & 3       &1.5                &68.4              &4.69            & 1.72                & 8.8             \\
  28                           &60     & NA & 3       &1.75              &65.9             &4.61            & 2.39               & 3.9                \\
  29                           &60     & NA  & 3       & 2             &61.5             &5.05           &2.57                & 1.1                 \\
  \enddata
  \tablecomments{$Z_{Fe},Z^{'}_{Fe},\frac{M_{disk}}{M_{2nd}}$ show the percent value without the percent sign ($\%$) here. The prime symbol indicates quantities of the post-encounter projectile. The close encounters are followed for $>$20 hours so that the projectile enters a new hyperbolic orbit ( $\delta x>20R_0$, see figure \ref{fig:encounter}) with a well formed debris disk (see figure \ref{fig:run8}).  The mass of the projectile's core is 1 Mercury core mass in all simulations initially. The radii of the projectiles range from $0.43R_0$ to $0.59R_0$. The core is exactly retained in most simulations and a maximum core lose of 2\% by mass occurs in run 27.}
\end{deluxetable}

In the most favourable case, when the spin always aligns with the orbital angular momentum in a series of close encounters, four encounters (similar to run 16) can raise $Z_{Fe}=30\%$ (chondritic value) to $Z_{Fe}\sim 70\%$ (present-day value). Encounters with spinning projectiles (we expect fast spinning projectiles except in the first encounter) show very small decrease of $v_\infty$ after the encounter (e.g., -0.11 km/s in run 16). \citet{Emsenhuber2019} shows a typical time interval about 0.1-1 Myrs between two successive hit-and-run collisions at 1 AU. They also found the median returning velocity is the same as the departing velocity when only the major planets are included. We expect the returning velocity to increase above the departing velocity during the 0.1-1 Myrs gravitational scattering with planetesimals in the early solar system. Such increment will be more pronounced when proto-Mercury's mantle is significantly removed after several encounters. A large returning speed will eventually help proto-Mercury to break out of the series of close encounters with proto-Venus.  More encounters are needed if the spin and orbital angular momentum happen to be anti-aligned leading to a low efficiency in mantle removal. 

The hit-and-run impact model \citep{Asphaug2014} also requires orbital crossing between proto-Mercury and proto-Venus. All single impact models require a high impact velocity around a specific impact angle \citep{Benz2007,Asphaug2014, Chau2018}. Multiple impacts/encounters are subject to the uncertainty of the returning collisions/encounters. We further note that the probability of having four consecutive close encounters ($d<2R_0$) without a collision ($d \sim 1.5R_0$) is only $\sim10 \%$ out of a simple geometric consideration. To assess the likelihood of these scenarios, hybrid N-body and hydrodynamical simulations \citep[see, e.g.,][]{Burger2019} are desirable to treat tidal interactions faithfully. It is noteworthy that our tidal disruption model does not involve intense shock heating so it can naturally explain the maintenance of moderate volatiles on present-day Mercury \citep{Peplowski2011}. The involved encounters can resurface the proto-Mercury and only the last encounter is expect to leave evidence on present-day Mercury for future space mission \citep[e.g. BepiColombo][]{Benkhoff2010}. In the meantime, the intruding silicate stream heats the deep mantle of proto-Venus (figure \ref{fig:run8} panel 4) due to shocks potentially leading to a superadiabatic entropy/temperature profile. This entropy profile contrasts with the subadiabatic entropy profile of proto-Earth after the Moon-forming giant impact \citep{Deng2019b}. The dichotomy in the initial conditions of Venus and Earth may provide an explanation for the different evolution path of Venus and Earth. We will present high-resolution simulations to study the thermal state of the post-encounter protoplanets in the future.

\section{Conclusions}
\label{sec:con}
We carried out hydrodynamical simulations for a series of hyperbolic encounters between proto-Mercury and proto-Venus which may occur due to dynamical instabilities in the early solar system. Such encounters can remove proto-Mercury's mantle, especially when its spin is aligned with the orbital angular momentum. In the most favourable case, four consecutive encounters can remove the mantle of a chondritic proto-Mercury to form the present-day iron-rich Mercury. We also found various outcomes of a close encounter, including tidal capture and formation of binary planets (likely to merge due to further tidal interactions), change of spin rates. These processes can dramatically affect simulations of the early solar system and should be taken into account as done for collisions in N-body simulations \citep{Chambers2013}. 

\bigskip

We thank the anonymous referee for constructive criticism that helped to improve the manuscript significantly. We acknowledge support from the Swiss National 
Science Foundation via a early postdoc mobility fellowship.

\software{GIZMO code \citep{Hopkins2015}, ballic \citep{Reinhardt2017}, VisIt}

\bibliographystyle{aasjournal}
\bibliography{references}




\end{document}